\documentclass[twoside]{ilcws08}
\usepackage[latin1]{inputenc}
\usepackage{graphicx}
\usepackage{epsfig}
\usepackage{color}
\usepackage{wrapfig}
\usepackage{amssymb,amsmath,array}

\begin{document}
%\title{Resolution Studies on fine Pitch Silicon Strip Sensors} %% 
\title{Silicon Detectors for the Large Prototype TPC test setup at DESY } %% 
%***********************************************************************
% AUTHORS INFORMATION AREA
%***********************************************************************
\author{S. Haensel$^1$, T. Bergauer$^1$, M. Dragicevic$^1$, J. Hrubec$^1$, M. Krammer$^1$, \\ A. Dierlamm$^2$, T. Barvich$^2$, F. Hartmann$^2$,  Th. Müller$^2$
% Optional short acknowledgment: remove next line if non-needed
%\thanks{This is an optional funding source acknowledgment.}
% DO NOT MODIFY THE FOLLOWING '\vspace' ARGUMENT
\vspace{.3cm}\\
% Addresses and institutions (remove "1- " in case of a single institution)
1 - Institute of High Energy Physics of the Austrian Academy of Sciences (HEPHY) \\
Nikolsdorfergasse 18, 1050 Vienna - Austria
%% Remove the next three lines in case of a single institution
\vspace{.1cm}\\
2 - Institut für Experimentelle Kernphysik - Forschungszentrum Karlsruhe (IEKP) \\
Hermann-von-Helmholtz-Platz 1, 76344 Eggenstein-Leopoldshafen - Germany\\
}
%%***********************************************************************
% END OF AUTHORS INFORMATION AREA
%***********************************************************************

\maketitle

\begin{abstract}

The Linear Collider TPC collaboration constructed a Large Prototype TPC (LPTPC) which is now installed at the EUDET facility, in DESY. The SiLC-collaboration (Silicon for the Linear Collider) will install position sensitive silicon strip sensors outside the field cage of the LPTPC, to provide precise tracking information. The data acquisition system (DAQ) is an adapted CMS readout system. The silicon modules are tested and ready to be installed, the mechanical module support and the DAQ system are in preparation.

\end{abstract}

\section{Introduction} 

The present paper describes the work of the SiLC-collaboration\cite{SILC} for the participation in the LPTPC test setup, with the purpose to participate at the LPTPC test setup at the EUDET facility in DESY. The SiLC-collaboration will design, build and install position sensitive silicon strip modules in the small gap between the LPTPC and the surrounding magnet complementing the data provided by the TPC. In the envisaged setup it will be possible to test modules with different sensors and readout chips. Furthermore it will allow to evaluate a first prototype of an external layer surrounding the TPC, as foreseen in the ILD concept, and to verify the corresponding simulations.

\section {The setup}

\begin{wrapfigure}[13]{r}{0.48\columnwidth}
\vspace{-45pt}
\centerline{
\includegraphics[width=0.5\columnwidth]{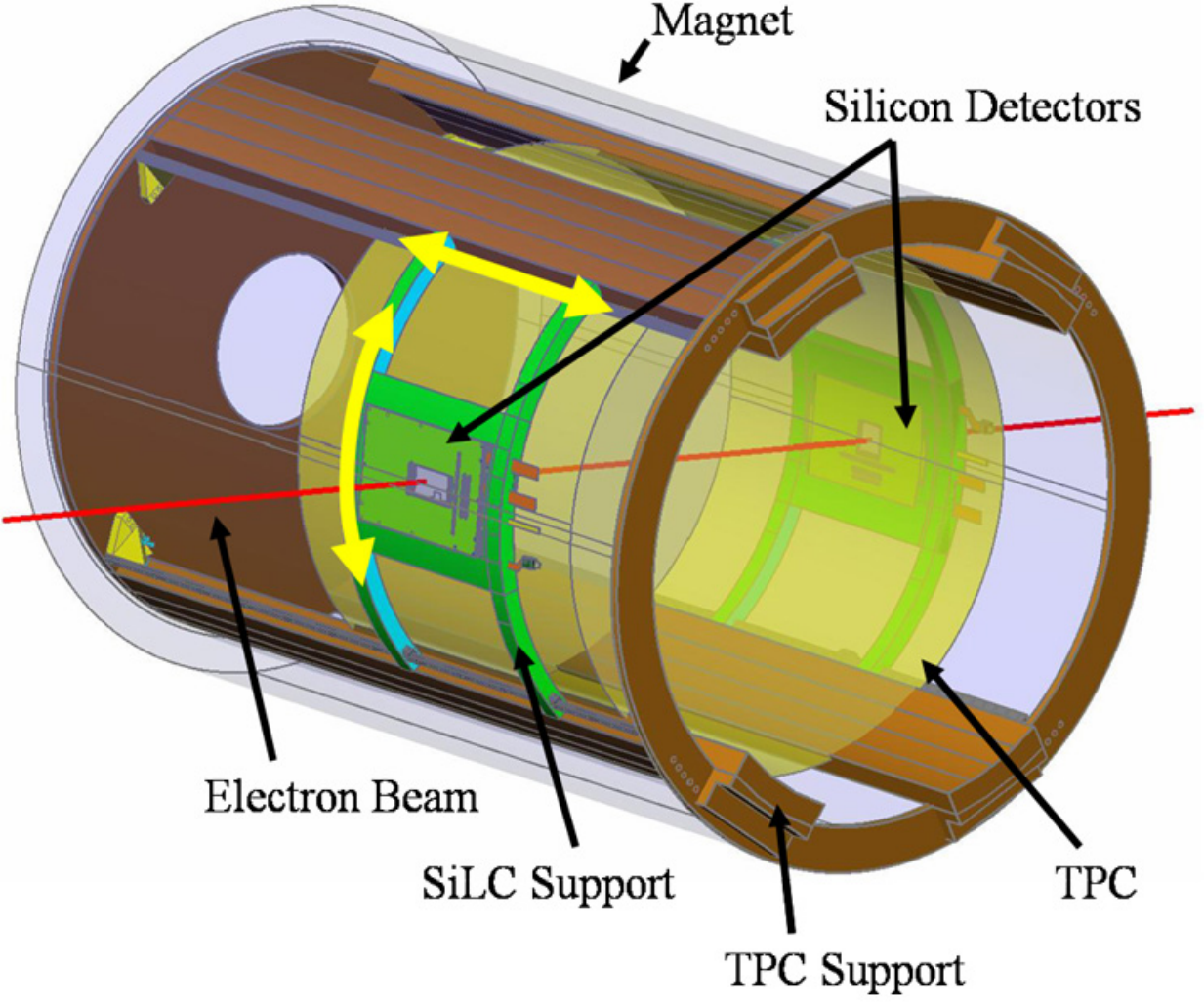}}
\vspace{-10pt}
\caption{LPTPC setup}
\label{Fig:setup}
\end{wrapfigure}

	The EUDET facility with the LPTPC test setup is located in a reserved DESY~II test beam area. Converted bremsstrahlungs-electrons from the accelerator will reach the test setup with an adjustable momentum between 1~and~6~GeV with a spread of about 2~\% and a divergence of 2~mrad. The estimated average rates for 6~GeV electrons are 250~Hz. The Large Prototype TPC with its support structure inside the magnet is shown in Fig.~\ref{Fig:setup}. The magnet is a Persistent Current, superconducting MAGnet (PCMAG), which will provide an inhomogeneous B-field of 1~Tesla. The space between the magnet and the TPC is just 35~mm in radius, thus leaving very little space for the silicon detectors and their appropriate support structure. The magnet can move in all directions in order to provide a scanning of the TPC. Hence the behaviour of the TPC can be investigated in the different regions of the inhomogeneous magnetic field.

An area of approximately 10~$\times$~20~mm$^2$ will be covered with silicon sensors, whereof in the beginning only a width of 38.4~mm is read out. This setup will simulate the part of the silicon envelope on either side of the TPC. In order to keep these two silicon modules aligned to the beam axis, the mechanical support structure of the silicon has to compensate for the movements of the magnet.

\section{Sensor tests}

\begin{wrapfigure}[18]{r}{0.48\columnwidth}
\vspace{-45pt}
\centerline{
\includegraphics[width=0.47\columnwidth]{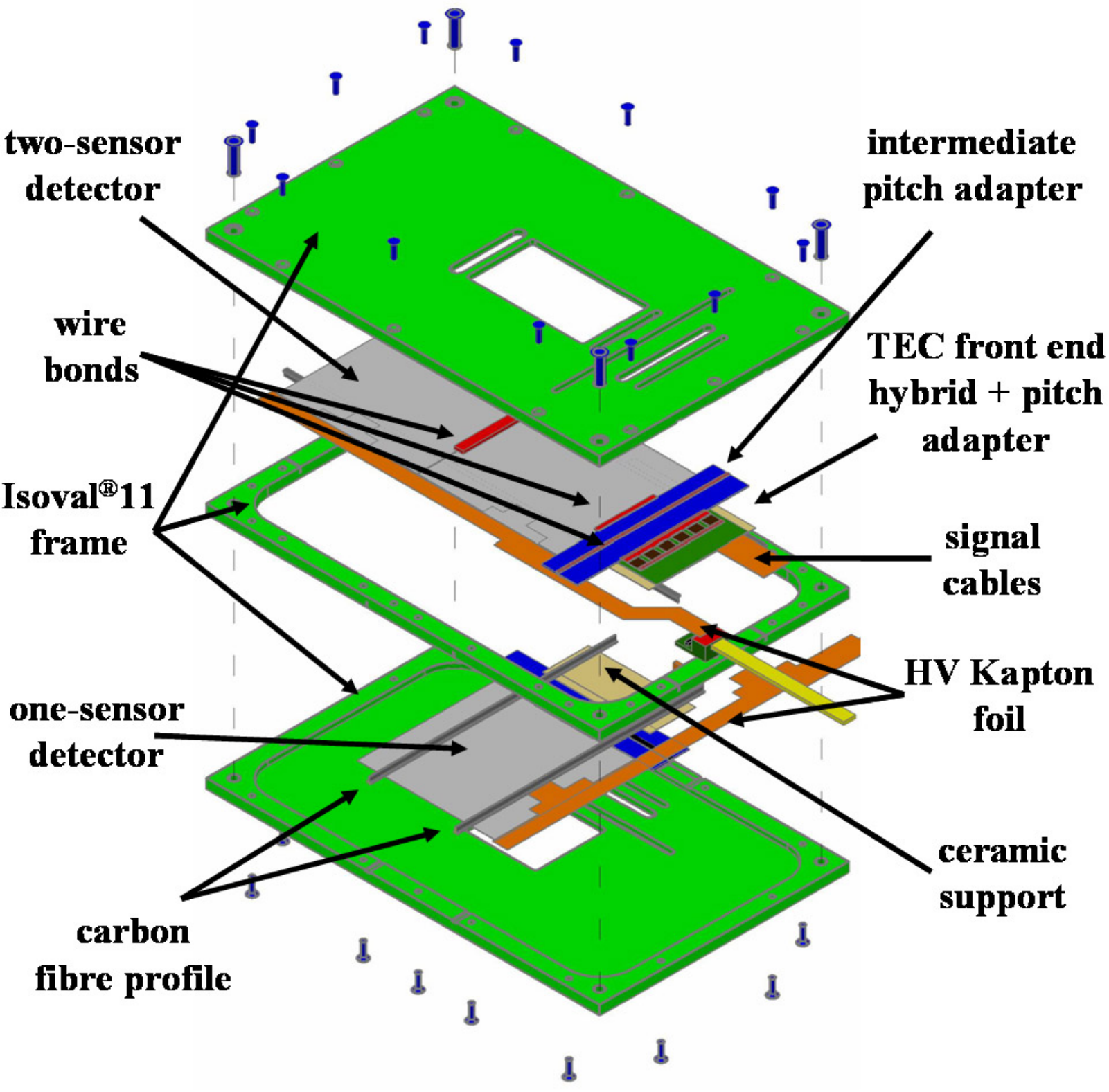}}
\vspace{-10pt}
\caption{Silicon module layout}
\label{Fig:mod}
\end{wrapfigure}

Each of the two detector modules consists of six silicon strip sensors. These single sided AC coupled silicon strip sensors with a size of 91.5 $\times$ 91.5 $\times$ 0.320~mm$^3$ were designed by the SiLC-collaboration and manufactured by Hamamatsu Photonics, Japan. Each sensor has 1792 readout strips with a readout pitch of 50~$\mu$m, without intermediate strips, biased with a 20~M$\Omega$ poly-silicon resistor. Prior to the module assembly, the sensors showed leakage currents below 300~nA up to an biasing voltage of 800~V and full sensor depletion at an average depletion voltage of 55~V. Single strip measurements revealed single strip currents of around 130~pA (at 100~V), values for the poly silicon resistors of above 20~M$\Omega$ and coupling capacitances in the order of 150~pF, indicating the excellent quality of the sensors.

\section{Detector modules}

\begin{wrapfigure}[11]{r}{0.48\columnwidth}
\vspace{-90pt}
\centerline{
\includegraphics[width=0.47\columnwidth]{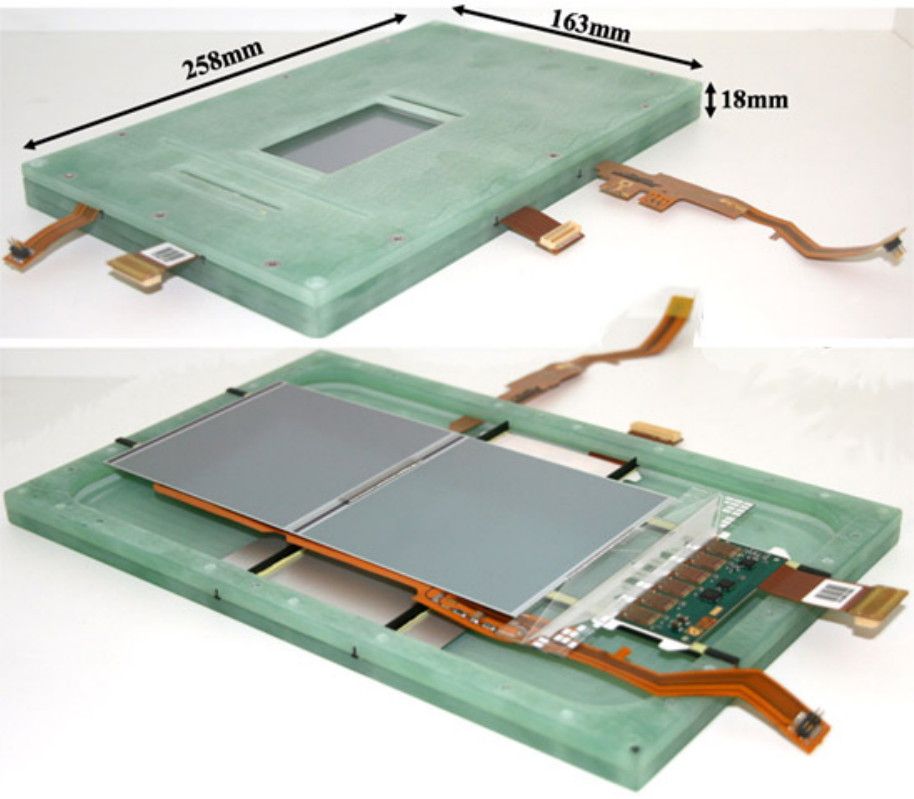}}
\vspace{-10pt}
\caption{Top: closed silicon module without light-tight cover; bottom: module without top layer of the frame}
\label{Fig:modulePic}
\end{wrapfigure}

The two detector modules (Fig.~\ref{Fig:mod}) each consist of a one-sensor and a two-sensor silicon detector and will be screwed onto the moveable support structure. They will be aligned in the beam axis in the small gap between the TPC and the magnet, one in front and one behind the TPC, to provide precise tracking information. The silicon sensors have a spatial resolution of 9~$\mu$m. This value was determined during a SiLC test beam at the SPS in 2008 \cite{Testbeam}. To provide this resolution in both directions (R$\varphi$ and z), the two silicon detectors in each module are assembled at an angle of 90°. The horizontal detector, consisting of two daisy-chained sensors, is located closer to the magnet than the other silicon detector, which is composed of one sensor. 

Each silicon detector consists of two carbon fibre T-beams as basic support. The sensor(s), the pitch adapter and the readout electronics are glued onto these T-profiles. In a following step of the module assembly these T-beams are clamped between three layers of the module frame (produced with Isoval11). Because new readout chips are not yet available, the first four detectors were constructed with front end hybrids from the CMS Tracker End Cap module production. Each of these hybrids contains 6 APV25 readout chips, providing a total of 768 readout channels, and can be read out by an adapted CMS readout system. It is not possible to use two of these hybrids side by side, therefore the sensitive silicon area per detector is reduced to 38.4~mm. The hybrids from the CMS production are connected to a pitch adapter designed to connect to a sensor with 143~$\mu$m pitch. A further intermediate pitch adapter was therefore necessary to adapt to the pitch of 50~$\mu$m of the used sensors. Kapton foils from the CMS module production are used to deliver the HV bias voltage to the sensor backplane and to isolate the sensor backplane from the carbon fibre profiles. Each HV-line contains a RC circuit to stabilize voltage fluctuations and to filter out high frequencies in the HV line. Fig.~\ref{Fig:modulePic} shows a finalised detector module, which has a thickness of 18~mm plus around 2~mm for a light-tight adhesive foil.

\begin{figure}[ht]
  \centering
  \vspace{-10pt}
    \includegraphics[width=0.85\textwidth]{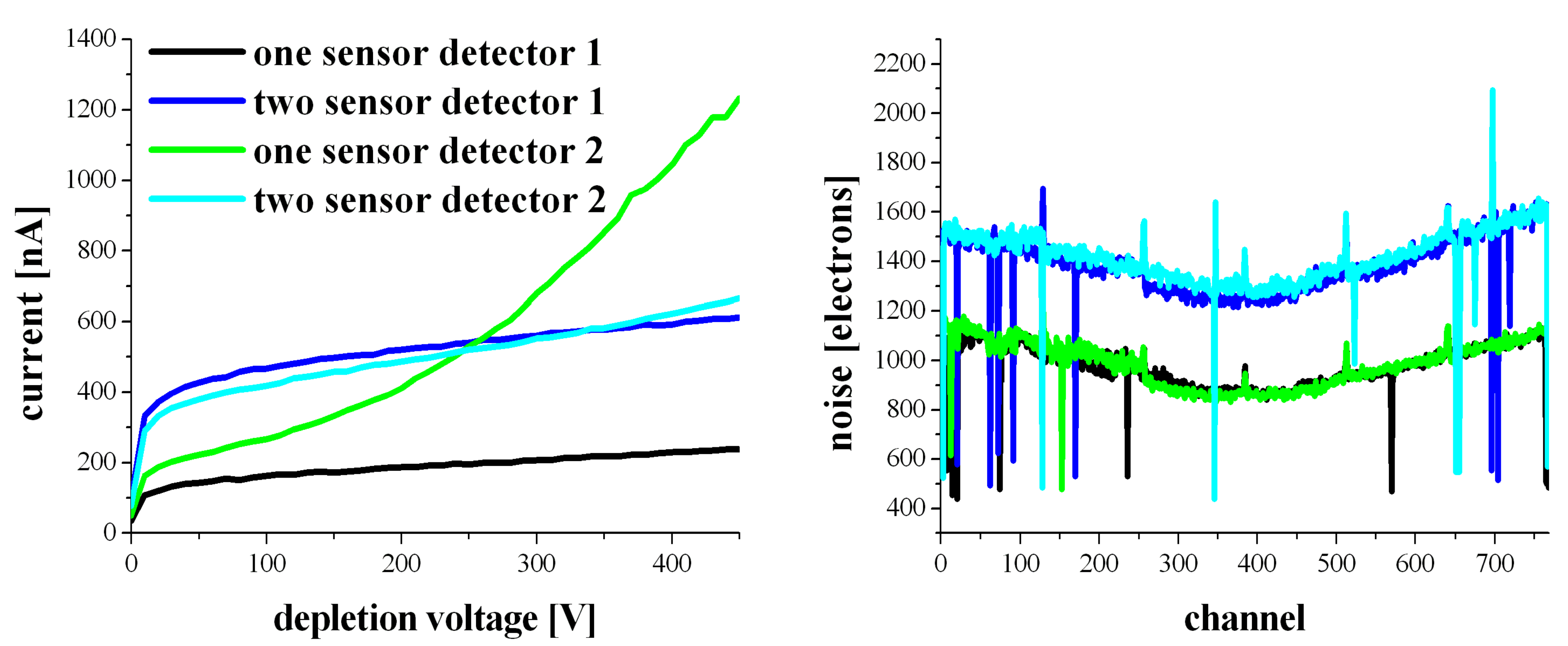}
\vspace{-10pt}
\caption{Electrical measurements of the silicon detectors - as expected, the detectors with two sensors have a higher current at 100~V and a higher noise than the detectors with one sensor}
\label{Fig:ARCtests}
\end{figure}

The two silicon modules were electronically tested: the leakage current of the silicon modules is below 1.5~$\mu$A at 450~V and the noise of the readout strips is as low as required (Fig. \ref{Fig:ARCtests}). In the noise plot a few channels with a very low noise indicate readout channels which are not connected to the readout chip. Investigations revealed defects on the intermediate pitch adapter. Since there are no clusters of disconnected readout channels in the central region of the sensors, the detector modules are well suited for this setup.

\vspace{-10pt}
\section{Mechanical support of the silicon modules}

The mechanical support has to position the two detector modules on each side of the TPC inside the gap of 35~mm between the magnet and the TPC. Since the sensitive areas of the modules have to be in the beam line, the support has to compensate vertical and horizontal movements and even horizontal rotations of the magnet.  

Design studies for this support are ongoing. The detector modules are constructed in a way that they leave a safety clearance of at least 4~mm to the TPC and the surrounding magnet. The R$\varphi$-position of the modules will be assigned by screwing them at different positions onto a curved sledge. This sledge will then be moved manually in and out of the magnet along two sliding rails, which will be mounted on the TPC support structure. 

The alignment of the silicon sensors and the TPC has to be performed with tracks that traverse the silicon on both sides of the TPC. First simulations are ongoing and with enough statistic and a homogeneous magnetic field it should be possible to achieve a relative alignment of the silicon and the TPC of a few $\mu$m \cite{Regler}.

\section{Data aquisition}

\begin{wrapfigure}[14]{r}{0.4\columnwidth}
\vspace{-40pt}
\centerline{
\includegraphics[width=0.39\columnwidth]{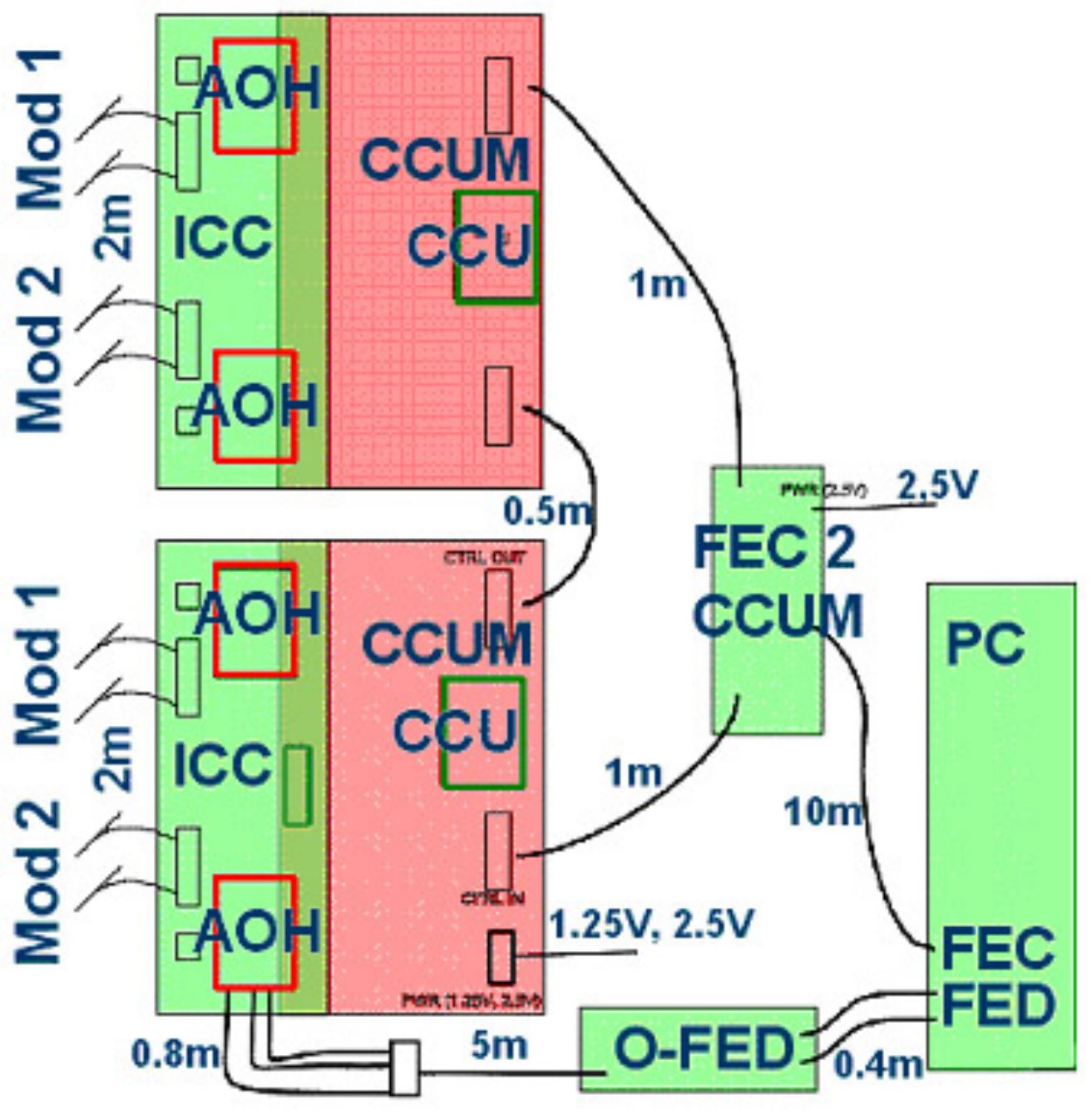}}
\vspace{-10pt}
\caption{Schematic of the adapted CMS DAQ}
\label{Fig:DAQ}
\end{wrapfigure}

For the first period an adapted CMS readout system will be used. Later on it is foreseen to use new developed readout chips from the SiLC collaboration with the appropriate DAQ. 

The adapted CMS readout was chosen because we have fully assembled front end hybrids from the CMS module production and IEKP has an adaptable CMS petal test readout system. Some parts have to be replaced, partly with new designed components. Fig.~\ref{Fig:DAQ} shows a sketch of the readout system that will be used in the LPTPC setup. 

The signals from the silicon sensors are read out by the APV25 chips on the CMS front end hybrids. Electrical cables bring the data to the Inter Connect Cards (ICC), sitting outside the magnet. There the electrical signals get converted to analogue optical signals in the Analogue OptoHybrids (AOHs). Via optical links these signals are transferred to the Optical Front End Driver (O-FED), re-converted to electrical signals, delivered to the FED card of our DAQ PC and stored. 

The readout is controlled from the DAQ PC via a Front End Control (FEC) card which steers the Central Control Units (CCU). The CCUs provide I$^2$C control sequence and clock via the ICC cards to the front end hybrids. To synchronise the TPC DAQ and the silicon DAQ a Trigger Logic Unit (TLU) will be used to centrally distribute a timestamp and the trigger signal, received from scintillators in front of the experiment.

\section{Summary}

Preparations to include the silicon detectors into the LPTPC test setup are progressing well. The design of the mechanical support has to be finalized before its construction will proceed. The silicon detector modules are already assembled and their functionality has been verified. In the first half of 2009 the adapted CMS readout system should be available and the silicon detector modules will be integrated in the LPTPC setup for combined data taking.

\section{Bibliography}

% ****************************************************************************
% BIBLIOGRAPHY AREA
% ****************************************************************************

\begin{footnotesize}
% IF YOU DO NOT USE BIBTEX, USE THE FOLLOWING SAMPLE SCHEME FOR THE REFERENCES
% ----------------------------------------------------------------------------

\end{footnotesize}

% ****************************************************************************
% END OF BIBLIOGRAPHY AREA
% ****************************************************************************

\end{document}